\documentclass[twocolumn,showpacs,aps,prl]{revtex4-1}
\usepackage{epsfig,amsmath,amssymb}

\newcommand{\sub}[1]{_{\mbox{\scriptsize{#1}}}}  

\begin{document}

\title{Localized collapse and revival of coherence in an ultracold Bose gas}
\author{J.~M. McGuirk and L.~F. Zajiczek}
\affiliation{Department of Physics\\
Simon Fraser University, Burnaby, British Columbia V5A 1S6, Canada}
\date{\today}

\begin{abstract}
We study the collapse and revival of coherence induced by dipolar spin waves in a trapped gas of $^{87}$Rb atoms.  In particular we observe spatially localized collapse and revival of Ramsey fringe contrast and show how the pattern of coherence depends on strength of the spin wave excitation.  We find that the spatial character of the coherence dynamics is incompatible with a simple model based only on position-space overlap of wave functions.  This phenomenon requires a full phase-space description of the atomic spin using a quantum Boltzmann transport equation, which highlights spin wave-induced coherent spin currents and the ensuing dynamics they drive.
\end{abstract}

\maketitle

Trapped ultracold gases are promising systems for compact, high resolution interferometric sensors.  They have been proposed for a diverse range of applications, including atomic clocks \cite{katori2005,deutsch2010}, magnetometers \cite{veng2007}, polarizability measurements \cite{sackett2008}, and inertial force sensors \cite{kasevich2009}.  Many interferometric sensors rely on differential energy shifts for measurements.  Furthermore, ultracold trapped gases all require inhomogeneous external potentials for confinement.  While this confinement allows for long interrogation times and high sensitivities, the inherent inhomogeneity can have profound impacts on the coherence of the sample.  Thus it is imperative to understand atomic coherence in trapped gases in the presence of inhomogeneous potential energies if they are to be used as interferometric sensors.

While Bose-Einstein condensates offer the potential for enhanced measurement sensitivity via squeezed states, it is often equally advantageous to use ultracold non-degenerate gases due to their larger populations, faster cycle times, and simpler experimental designs \cite{cronin2009}.  Even above degeneracy, however, gases can exhibit quantum scattering effects if the de Broglie wavelength of the atoms is long compared to the scattering length $a$. In this case, interferometric measurements using non-degenerate quantum gases must contend with the identical spin rotation effect (ISRE) \cite{isre}.  Exchange symmetry between identical particles in a quantum gas leads to interference effects: if the interacting particles have different spins, a net spin rotation can occur.  This spin rotation gives rise to a wide array of macroscopic collective behaviors, including spin-state segregation \cite{lewando2002,thomas2008}, spin waves \cite{mcguirk2002, mcguirk2010}, spin locking in coupled fluids \cite{mcguirk2003}, and dramatically enhanced coherence times \cite{deutsch2010}.

Of particular concern for trapped atom-based measurements is the collapse and revival of coherence, which was first noted in \cite{williams2002,laloe2002,levitov2002}.  A recent study of the collapse and revival of interference in the time domain showed that the dynamics could be tuned by altering the differential potential experienced by the two-state system \cite{deutsch2010}.  If the trap frequencies are larger than all other frequency scales, especially the collision rate and differential frequency shifts, as in \cite{deutsch2010}, the ensemble remains largely spin-synchronized.  (See \cite{laloe2009} for a detailed discussion of time scales.) This limit obscures the spatial character of the spinor evolution, and for experiments outside this limit, a more detailed approach is required.

In this Letter, we observe localized collapse and revivals of interference fringes and demonstrate how the spatial character of the localization changes with excitation strength.  Furthermore, we show that a simple, intuitive model based on the degree of spatial overlap of the spin-state wave functions fails to explain the observed interference patterns, requiring a full numerical treatment using a quantum Boltzmann equation for spin transport.

We interferometrically study the coherent evolution of atomic spins in the presence of spin wave excitations.  The spin system described herein consists of two magnetically trapped hyperfine states of $^{87}$Rb, $|1\rangle = |F=1, m_F=-1\rangle$ and $|2\rangle = |F=2, m_F=1\rangle$, which form a pseudospin-1/2 doublet.  A two-photon microwave transition couples $|1\rangle$ and $|2\rangle$ and creates an equal coherent superposition.  We excite spin waves in this superposition with an inhomogeneous differential optical potential $U\sub{diff}$ designed to drive the lowest normal spin wave mode: the dipole mode \cite{mcguirk2010}.

The optical potential consists of a diode laser detuned by 0.3~nm from the $D_2$ cooling transition, and the slight difference in the energies of the hyperfine ground states gives slightly different detunings for each state, producing a differential AC Stark shift.  A frequency and amplitude modulated acousto-optic modulator sweeps the laser across the atom cloud and creates a differential potential energy that varies linearly with axial position $z$ (see \cite{mcguirk2010}).  The gradient of the spin perturbation, $dU\sub{diff}/dz$, is controlled by adjusting the laser power.  The Zeeman shift from the trapping potential and the mean-field shift also contribute to the differential potential, but these sources of inhomogeneity can be eliminated with a mutual compensation scheme by tuning the magnetic field \cite{lewando2002}.

Typical experimental parameters include a magnetic field of $B_0 = 3.05$~G at the center of the trap for this mean-field compensation, with a peak density of $n_0 = 2.5 \times 10^{19}$~m$^{-3}$ and temperature of 650~nK, about 60\% above the critical temperature for condensation.  The cylindrically symmetric magnetic trap is highly elongated in the axial direction with an aspect ration of 37:1, corresponding to trap frequencies of 247 and 6.7~Hz.  Thus, the density may be radially averaged, and all dynamics can be treated one-dimensionally.

\begin{figure}[!btH]
\leavevmode
\epsfxsize=3.375in
\epsffile{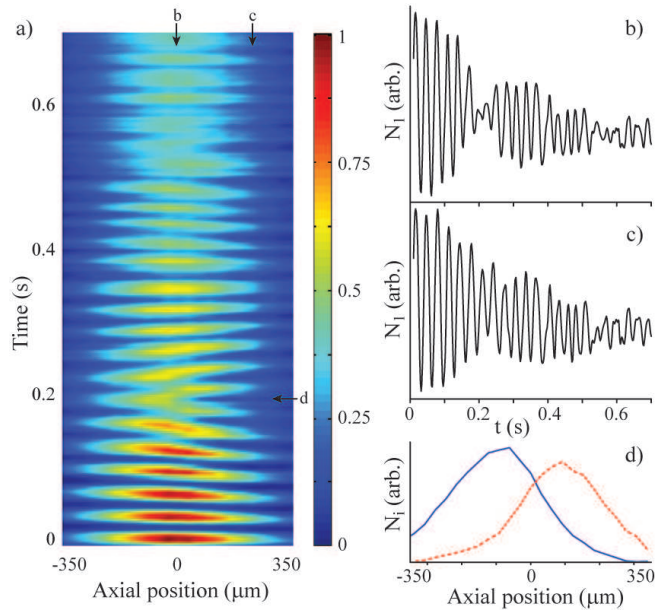}
\caption{\label{fig:fringes} (color online) Ramsey fringes in the presence of a dipolar spin wave driven by an optical potential with $dU\sub{diff}/dz \simeq h\times10$~Hz/mm.  (a)  $N_1$ measured after a Ramsey sequence, radially averaged and normalized to its maximum value, and plotted in false color as a function of $t$ and $z$. Data is interpolated between bins and time steps.  (b)-(c) Cross sections of Ramsey fringes for the center bin and a bin near the edge of the cloud respectively, as shown by the arrows in (a).  (d) Normalized projections of the of the states of the superposition $N_1$ (solid line) and $N_2$ (dashed line) at the time of maximal component separation and collapse, denoted by an arrow in (a).}
\end{figure}

Ramsey interferometry allows us to measure the local transverse components of the atomic spin using two $\pi/2$-pulses separated by a variable delay time $t$.  Following the interferometer sequence, absorption imaging measures the number of atoms $N_1$ returning to the $|1\rangle$ state.  The microwave coupling is detuned a small amount from the  $|1\rangle-|2\rangle$ transition ($\delta\sim30$~Hz), and as we vary $t$ we observe high contrast interference fringes (Fig.~\ref{fig:fringes}).  Alternately we can omit the final $\pi/2$-pulse to project the superposition onto the $|1\rangle$ state to measure $N_1$, or replace the final $\pi/2$-pulse with a $\pi$-pulse to measure $N_2$.  For all measurements, we measure the atomic distributions with absorption imaging and bin the images into equally spaced axial bins.

 The optical potential excites a dipole spin wave mode that causes a linear phase gradient to build up across the cloud, as seen by the tilting of the phase fronts in Fig.~\ref{fig:fringes}(a).  ISRE-induced spin currents build up and lead to longitudinal rotation of the spin.  This spin rotation manifests as axial separation of the spin states in the magnetic trap [Fig.~\ref{fig:fringes}(d)], and as the spin wave peaks, the phase fronts become disrupted and interference fringe visibility collapses.  The spins soon rephase, and the spin oscillation continues, with a second collapse occurring at the next peak in the spin wave cycle.

There is an overall damping of both the spin wave excitation and the superposition coherence through elastic collisions that eventually randomize the atomic spins and destroy the coherent superposition.  This damping is more pronounced in potentials with greater inhomogeneity.  There is also longitudinal relaxation due to dipolar collisions leading to loss from the $|2\rangle$ state.  Lastly the spin waves themselves damp by an additional mechanism: Landau damping due to coupling to higher order, highly damped spin wave modes \cite{mcguirk2002,williams2002,laloe2002}.

\begin{figure}[!tbH]
\leavevmode
\epsfxsize=3.375in
\epsffile{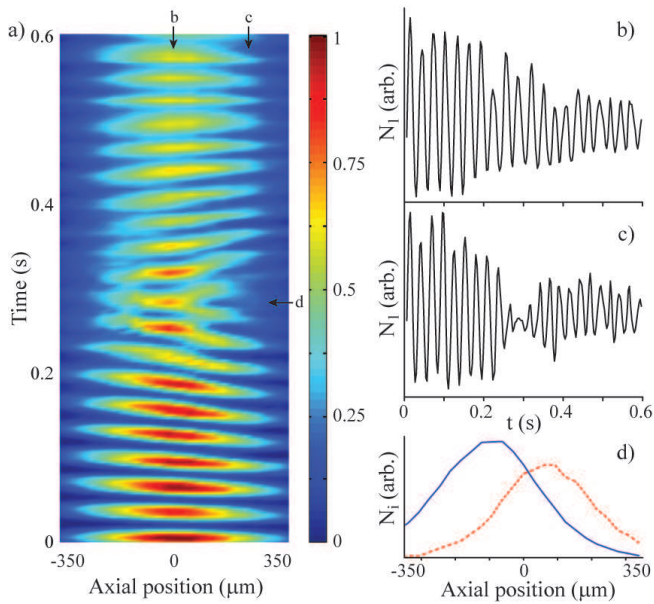}
\caption{\label{fig:fringes2} (color online) Ramsey fringes in the presence of a more weakly driven dipolar spin wave ($dU\sub{diff}/dz \simeq h\times7$~Hz/mm).  (a)  $N_1$ measured after a Ramsey sequence, normalized to its maximum value.  (b)-(c) Cross sections of Ramsey fringes for the center bin and a bin near the edge of the cloud respectively, as shown by the arrows in (a).  (d) Normalized projections of $N_1$ (solid line) and $N_2$ (dashed line) at the time of maximal component separation and collapse, denoted by an arrow in (a).}
\end{figure}

The spatial pattern of the interference contrast is particularly interesting, as an ensemble-wide collapse does not occur.  There is nearly complete collapse in the center of the atom cloud [Fig.~\ref{fig:fringes}(b)], but in the wings of the distribution, fringe visibility is only moderately affected at the peak of the spin wave perturbation [Fig.~\ref{fig:fringes}(c)].  Furthermore, this spatial pattern of interference is not identical for all spin wave-induced collapses.  Figure~\ref{fig:fringes2} displays Ramsey fringes for a more weakly driven spin wave excitation.  When the fringe collapse occurs, it happens to a greater degree near the edges of the cloud and not as much in the center [Figs.~\ref{fig:fringes2}(c) and (b) respectively].  Note that the time scales are different, as spin wave frequency has a nonlinear dependence on excitation amplitude in this system \cite{mcguirk2010}.

\begin{figure}[!tbH]
\leavevmode
\epsfxsize=3.375in
\epsffile{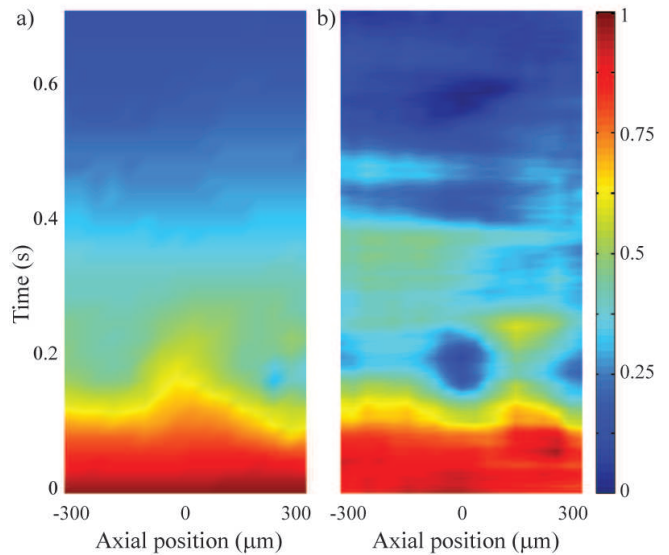}
\caption{\label{fig:contrast} (color online) (a) Predicted interference contrast $F\sub{o}$ determined from population overlap (see text). (b) Measured fringe contrast, showing minima in the cloud center at $t\simeq180$ and 575~ms, calculated by performing Fourier transforms on the Ramsey fringes in Fig.~\ref{fig:fringes}(a) using a two-fringe running window to extract the fringe amplitude.  The resulting contrast is filtered to remove residual aliasing at the Ramsey frequency.}
\end{figure}

To highlight the need for a complete phase-space analysis of the spin dynamics, we construct a simple model for fringe visibility $F\sub{o}$ based on the spatial overlap of the two spin states in the superposition.  The local relative populations of the superposition, $N_1(z)$ and $N_2(z)$, are projected as described previously [e.g. Fig.~\ref{fig:fringes}(d)].  If the interference contrast were determined only by the local density overlap, then it would be given by
\begin{equation}
F\sub{o}(t,z) = \frac{2\sqrt{N_1(t,z)N_2(t,z)}}{N_1(t,z)+N_2(t,z)} e^{-t/\tau},
\end{equation}
where $\tau$ is an empirically determined time constant that characterizes the cloud-wide decoherence due to elastic scattering.  We set $\tau\simeq350$~ms using an average of the decoherence times over the entire distribution.  Figure~\ref{fig:contrast}(a) shows how $F\sub{o}$ evolves in the presence of the same dipolar spin wave perturbation shown in Fig.~\ref{fig:fringes}, while Fig.~\ref{fig:contrast}(b) shows the actual measured fringe contrast.  The measured contrast is more sensitive to shot-to-shot magnetic field-induced phase noise, and the reduced contrast at $\sim150$~ms on the positive side of the trap (where the $|2\rangle$ state density is highest) is caused by atom loss due to dipolar relaxation.

The large component separation means the longitudinal rotation of the spin is especially pronounced on the edges of the cloud during the spin wave peak ($t\sim180$~ms), and thus the density overlap method predicts the smallest fringe contrast to be there.  In the center of the atomic distribution the density overlap is maximal, and one would expect higher contrast than on the edges [Fig.~\ref{fig:contrast}(a)].  This prediction clearly does not explain the actual collapse and revival dynamics observed in the strong dipole potential, which gives minimal contrast in the cloud center during the spin wave peaks [Fig.~\ref{fig:contrast}(b)].  While the local density overlap does give a maximum possible value for interference contrast at a given position, it provides an incomplete picture.

We turn to a full phase-space treatment of the spin vector to understand the spin dynamics completely.  The spin evolution can be described with a quantum Boltzmann transport equation for the spin momentum and position distribution $\vec{\sigma} (p,z,t)$ \cite{williams2002,laloe2002,levitov2002}:
\begin{equation} \label{boltzmann}
\frac{\partial \vec{\sigma}}{\partial t} + \frac{p}{m}\frac{\partial \vec{\sigma}}{\partial z} - \frac{\partial U\sub{ext}}{\partial z} \frac{\partial \vec{\sigma}}{\partial p} -
    \vec{\Omega} \times \vec{\sigma} = \frac{\partial \vec{\sigma}}{\partial t}{\Big |}\sub{1D},
\end{equation}
where $m$ is the atomic mass, $U\sub{ext}$ is the axial harmonic trapping potential, and the right hand side represents a collisional relaxation term related to the elastic scattering rate.  The differential potential $U\sub{diff}$ and mean-field coupling $g$ that drive spin waves are contained in the cross term  $\vec{\Omega} \times \vec{\sigma}$ as follows,
 \begin{equation}
\vec{\Omega} = ( g\vec{S} + U\sub{diff}\,\hat{w}) /\hbar.
\end{equation}
The mean-field coupling constant is $g=4\pi \hbar^2 a /m$, and $\hat{w}$ is a longitudinal unit vector on the Bloch sphere.  The spatial spin distribution is obtained by integrating $\vec{\sigma}$ over the momentum distribution, $\vec{S}(z) = \int \vec{\sigma}dp$.

\begin{figure*}[!tbH]
\leavevmode
\epsfxsize=7in
\epsffile{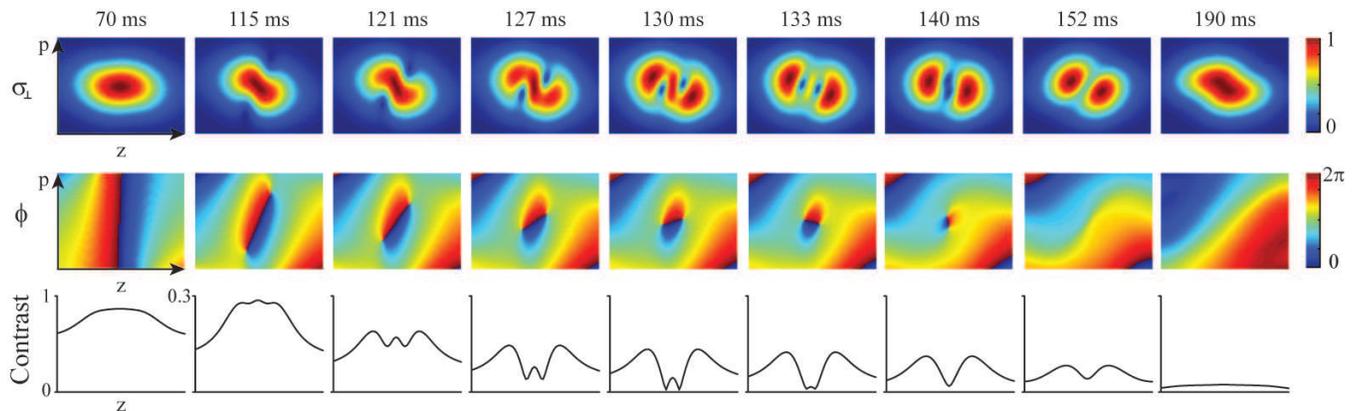}
\caption{\label{fig:phasespace} (color online) False color images from numerical simulations of the evolution of $\sigma_\perp$ (top) and $\phi$ (middle) in phase space under the influence of a dipolar spin wave, using similar parameters as for the data shown in Fig.~\ref{fig:fringes}.  This initial distribution at $t=0$ is Gaussian in $p$ and $z$.  At each time step $\sigma_\perp$ is normalized and does not reflect the global relaxation of the spin vector.  Bottom row: interference fringe contrast (transverse spin distribution $S_\perp$) obtained by integrating $\vec{\sigma}_\perp$ over the momentum coordinate.  The vertical scale for $t>70$~ms has been magnified by a factor of 3 to compensate for transverse spin relaxation.}
\end{figure*}

We numerically solve Eq.~\ref{boltzmann} using an alternative direction implicit finite difference method.  The transverse component $\vec{\sigma}_\perp$ describes the coherence.  Figure~\ref{fig:phasespace} shows the magnitude of $\vec{\sigma}_\perp$ and its orientation $\phi$ for a selection of evolution times, as well as the Ramsey fringe contrast obtained by integrating $\vec{\sigma}_\perp$ over $p$ to find $|\vec{S}_\perp|$.

The full phase-space evolution of $\vec{\sigma}_\perp$ sheds light on the spatial nature of the coherence as follows.  The effect of $U\sub{diff}$ manifests in Fig.~\ref{fig:phasespace} as an axial phase gradient (vertical phase stripes) for $t<100$~ms.  As the atoms move in the confining potential, these phase gradients correlate with momentum and position, and coherent spin currents develop on either side of the trap ($t\sim100-120$~ms).  The ISRE keeps the phase within each lobe of the spin distribution more uniform than the differential potential would otherwise suggest. The two lobes have opposite phases from each other, however, and where they meet, nodes form in the spin distribution.  When the counterpropagating spin currents meet at the peak of the spin wave cycle, exchange scattering rephases the two halves of the distribution ($t\sim130-140$~ms, corresponding to maximum spin-state segregation).  If the phases of the two lobes are significantly different, this rephasing leads to a zero in the center of the transverse spin distribution and manifests as a localized collapse of coherence.  Coherence revives somewhat as the spin wave completes a cycle ($t > 150$~ms) due to the ISRE rephasing the entire spin distribution, although elastic scattering destroys the superposition on a similar time scale.

For smaller amplitude spin wave excitations, the nodes develop more slowly and do not reach the center by the time the spin wave peaks.  Instead they are ejected from the phase-space profile, and thus the regions of minimal contrast do not reach the center of the distribution, as in Fig.~\ref{fig:fringes2}.  Conversely, if the spin wave amplitude is increased further, phase accrues more rapidly, and more nodes begin to appear.  The presence of contrast minima in the wings of the distribution at $t\simeq180$~ms in Fig.~\ref{fig:contrast}(b) may indicate that we are beginning to enter this regime.  Decoherence from elastic scattering increases as well for more strongly driven spin waves, and thus detection of more complex spatial features is difficult.

We have demonstrated that the coherence of trapped gas interferometers near quantum degeneracy can exhibit complex spatial patterns of collapse and revival.  This behavior can be understood by considering the full phase space evolution of the spin distribution given by a quantum Boltzmann equation.  Although we cannot tomographically reconstruct the full $\vec{\sigma}$ distribution, the effects of quantum dynamics in the momentum coordinate are still manifested in the spatiotemporal coherence patterns.  Clearly care must be taken to avoid significantly inhomogeneous differential potentials when attempting to make interferometric measurements.  Such inhomogeneous potentials enhance dephasing, and even worse, can lead to complete collapse of coherence.  If inhomogeneous potentials are unavoidable, their effect may be lessened by working with tight confining potentials and lower densities, as in \cite{deutsch2010}, where the entire atomic distribution tends to stay phase synchronized to a greater extent.

We thank Malcolm Kennett for helpful discussions.  This work was supported by NSERC and the CFI, and L.Z. acknowledges support from the SFU-VPR Fellowship Program.

\end{document}